# Does infrastructure investment lead to economic growth or economic fragility? Evidence from China

Atif Ansar,\* Bent Flyvbjerg,\*\* Alexander Budzier,\*\*\* and Daniel Lunn\*\*\*\*

**Abstract:** China's three-decade infrastructure investment boom shows few signs of abating. Is China's economic growth a consequence of its purposeful investment? Is China a prodigy in delivering infrastructure from which rich democracies could learn? The prevalent view in economics literature and policies derived from it is that a high level of infrastructure investment is a precursor to economic growth. China is especially held up as a model to emulate. Politicians in rich democracies display awe and envy of the scale of infrastructure Chinese leaders are able to build. Based on the largest dataset of its kind, this paper punctures the twin myths that (i) infrastructure creates economic value, and that (ii) China has a distinct advantage in its delivery. Far from being an engine of economic growth, the typical infrastructure investment fails to deliver a positive risk-adjusted return. Moreover, China's track record in delivering infrastructure is no better than that of rich democracies. Investing in unproductive projects results initially in a boom, as long as construction is ongoing, followed by a bust, when forecasted benefits fail to materialize and projects therefore become a drag on the economy. Where investments are debt-financed, overinvesting in unproductive projects results in the build-up of debt, monetary expansion, instability in financial markets, and economic fragility, exactly as we see in China today. We conclude that poorly managed infrastructure investments are a main explanation of surfacing economic and financial problems in China. We predict that, unless China shifts to a lower level of higher-quality infrastructure investments, the country is headed for an infrastructure-led national financial and economic crisis, which is likely also to be a crisis for the international economy. China's infrastructure investment model is not one to follow for other countries but one to avoid.

\* Saïd Business School, University of Oxford, e-mail: atif.ansar@sbs.ox.ac.uk
\*\* Saïd Business School, University of Oxford, e-mail: bent.flyvbjerg@sbs.ox.ac.uk
\*\*\* Saïd Business School, University of Oxford, e-mail: alexander.budzier@sbs.ox.ac.uk
\*\*\*\* Department of Statistics, University of Oxford, e-mail: d.lunn@stats.ox.ac.uk

This work was supported by funding from BT Group plc via the Saïd Business School, University of Oxford. We thank Ariell Ahearn, Benjamin Craddock, Philipp Dreyer, and Denis Tenchurin for their research assistance. We are grateful to Messrs Christopher Bennett, John Besant-Jones, Richard Bullock, Zhao Gaungbin, John Lee, and Gregory Wood for advice on data collection. The authors also wish to thank Professors Gordon Clark, Simon Cowan, Jim Hall, Dieter Helm, Colin Mayer, Roger Vickerman, and participants of the Infrastructure Transitions Research Consortium (ITRC) at St Catharine's College, Cambridge (March, 2014) for their comments on earlier drafts of the paper.
doi:10.1093/oxrep/grw022

## I. Introduction

Only if infrastructure investment 'grows by 15 to 18 percent (per year), can we reach 8 percent economic growth' said Mr Zeng Peiyan, the former minister in charge of China's State Development Planning Commission[1] (*The New York Times*, 24 September 1998). At the time, Asia was in the midst of a financial crisis. Redoubling investment in infrastructure was China's strategy to slip past the regional downturn. Mr Peiyan's view finds emphatic support in the extant literature in economics and with policy experts. A larger stock of infrastructure is thought to fuel economic growth by reducing the cost of production and transport of goods and services, increasing the productivity of input factors, creating indirect positive externalities, and smoothing the business cycle.

Using the case of China, this article explores a salient paradox in the theory on infrastructure. The macro-level school of thought that has dominated the mainstream discourse in economics has argued that increased public-sector investment in infrastructure (particularly in transport) 'increases the efficiency and profitability of the business sector; [and] this increase stimulates business investment in private capital (Aschauer, 1989*a*; 1989*b*)' in Banister and Berechman (2000, p. 134). In contrast, micro-level evidence from case studies and large datasets, typically published in planning and management journals, has shown that the financial, social, and environmental performance of infrastructure investments is, in fact, strikingly poor (Flyvbjerg *et al.*, 2002, 2003, 2005, 2009; Flyvbjerg and Budzier, 2011; Ansar *et al.*, 2014). The public sector is not uniquely challenged. Private firms also systematically bungle big capital investments (Nutt, 1999, 2002; Titman *et al.*, 2004; Flyvbjerg and Budzier, 2011; Ansar *et al.*, 2013; Van Oorschot *et al.*, 2013). The cost overrun and benefit shortfall on the Channel tunnel were so large that Anguera (2006, p. 291) concluded, 'the British Economy would have been better off had the Tunnel never been constructed'. The Danish Great Belt rail tunnel proved financially non-viable even before it opened (Flyvbjerg, 2009). Similarly, based on the largest dataset of its kind on the outcomes of 245 large dams, Ansar *et al.* (2014) found that the capital sunk into building nearly half the dams could not be recovered. How can poor outcomes of individual infrastructure investment projects amount to economic welfare in the aggregate? The macro and micro studies seem at loggerheads over the impact of infrastructure investments on economic prosperity.

In tackling the macro versus micro paradox of infrastructure investment, this article focuses on two research questions: (i) what are the outcomes of specific infrastructure projects in China, particularly in terms of cost, time, and benefit performance? and (ii) how do micro-level project outcomes link with macro-level economic performance? We selected China for the following two reasons.

First, given its high infrastructure investment and economic growth, China seems to fit the macro-level theories. For example, Démurger (2001) and Banerjee *et al.* (2009) argue that investment in and proximity to transport infrastructure have had a positive effect on economic growth in Chinese cities and provinces. But if even the data from China were found not to fit the macro-level theories, then it would call into question the fundamental soundness of the conventional wisdom. When it comes to testing competing theories of infrastructure, China's experience is a critical case.

---

[1] The State Planning Commission and the State Development Planning Commission are predecessors to what is now known as the National Development and Reform Commission of the People's Republic of China (NDRC). NDRC is a powerful macroeconomic management agency under the Chinese State Council.

Second, many scholarly articles, think tank reports, and the media see China as particularly effective in infrastructure delivery (Friedmann, 2005; Wolf, 2006; *The Economist*, 14 February 2008; Newman, 2011; McKinsey, 2013). The scale and speed of China's delivery of a massive stock of infrastructure since the mid-1980s evoke awe; democracies, in contrast, live in the 'slow lane' claims *The Economist* (2011).

Despite the widespread admiration of China's infrastructure development, there is scant bottom-up evidence from the field about the actual outcomes of specific investment projects. The macroeconomic account of infrastructure investments in China, for instance, omits the massive costs incurred in the building of megaprojects. Even proponents of more infrastructure in China, Banerjee *et al.* (2009, p. 5), acknowledge: 'We cannot use our results to estimate the social or private return on investing in transport infrastructure because we have no idea of the relevant costs.' They continue, 'Public investment in infrastructure may . . . be desirable, though . . . we would need cost data to be able to speak definitively about that' (p. 6). We thus aim to resolve this precise shortcoming of the extant literature in this article. Specifically, we report results on 95 road and rail transport infrastructure projects built in China from 1984 to 2008 and comparative results with a dataset of 806 transport projects built in rich democracies. In doing so, we also offer a more generalizable perspective on problems associated with managing major infrastructure projects and their consequences on the wealth (or poverty) of nations.

Transport infrastructure is an apt setting because conventional economic theory typically treats all infrastructure as an exogenous cost-reducing technological input into the economy, reflected via the proxy of *transport costs* (Krugman, 1991; Holtz-Eakin and Lovely, 1996; Glaeser and Kohlhase, 2004). For instance, stylized models in 'new economic geography' *à la* Krugman treat economic decisions regarding location of production resting heavily on 'costs of moving goods over space' (Glaeser, 2010, p. 7). The seemingly intuitive assumption in these models is that more and better infrastructure reduces the cost of transporting goods and services. The origin of this assumption is Paul Samuelson's concept of 'ice-berg' transport costs—i.e. 'one assumes that a part of goods "melts" during the transport between one region to the other' (Charlot, 2000, p. 2).

The results reported here challenge the traditional macro view. The evidence suggests that poor project-level outcomes translate into substantial macroeconomic risks: accumulating debt and non-performing loans; distortionary monetary expansion; and lost alternative investment opportunities. We hypothesize that debt-financed overinvestment in infrastructure contributes to underperformance and instability in the economy. Finally, we advance policy propositions grounded in our findings to enable policy-makers in China and elsewhere to improve the quality of decisions pertaining to infrastructure investments.

## II. Macro view of infrastructure and growth

The study of infrastructure investment in economics has been prone to recurring and then fading 'speculative bubbles of economics research' (Gramlich, 1994, p. 1176).

David Aschauer (1989*a*,*b*,*c*, 1993) triggered the most recent of these bursts of activity, particularly in the empirical literature. His series of papers sought to establish an econometric link between macro-level infrastructure investment and aggregate productivity. Paul Krugman's (1991) theoretical model, in which infrastructures such as road and rail lowered transport costs enabling increasing returns, strengthened the intuition underpinning Aschauer's empirical work. Alicia Munnell's work (1990, 1992) buttressed Aschauer's findings. Munnell (1990, p. 70) argued:

> The conclusion is that those [US] states that have invested more in infrastructure tend to have greater output, more private investment, and more employment growth. This evidence supports results found in earlier studies. The empirical work also seems to indicate that public investment comes before the pickup in economic activity and serves as a base, but much more work is required to spell out the specifics of the link between public capital and economic performance.

Aschauer and Munnell's macro-level studies and the 'new economic geography' *à la* Krugman set the tone for a slew of publications in the next two decades in academic journals in economics that, notwithstanding their nuances, advanced the primary claim that more public investment in infrastructure is better. Sanchez-Robles (1998, p. 106), for example, found 'a positive impact of public capital on the growth rate of output during the transition to a steady state' in two different samples of countries. Fernald (1999) found that the US interstate highway system was highly productive. Vehicle-intensive industries benefited from road building in particular.

Similarly, Fan and Zhang (2004) and Donaldson (2010) advanced the proposition that infrastructure supported increased income and productivity: Using data on rural infrastructure, Fan and Zhang (2004, p. 213) found that:

> [First] investing more in rural infrastructure is key to an increase in overall income of the rural population. Second, the lower productivity in the western region is explained by its lower level of rural infrastructure, education, and science and technology.

They proposed increasing the level of public capital 'to narrow' the difference in productivity between poorer regions and other regions. Similarly, using observations on trade flow data between 45 regions in India, Donaldson (2010, p. 1) advocated that more investment in railroads 'reduced trade costs, reduced interregional price gaps, and increased trade flows'.

Despite its broad appeal, the Aschauer and Munnell line of thinking was not universally accepted even among other macro scholars. A series of papers—e.g. Eisner (1991); Gramlich (1994); Evans and Karras (1994); Holtz-Eakin and Schwartz (1995)—though generally sympathetic to the basic argument, brought into question the research design, methods, and the robustness of causal inference of the Aschauer-style studies. Instead of overturning the results of the earlier studies, macro scholars took the discussion in a different direction. Where direct productivity effects were found to be weak or not found at all, the macro-studies began to insist on indirect impacts through spillover effects. Using aggregate and regional-level data from Spain, Pereira and Roca-Sagalés (2003, p. 238), for example, argued:

aggregate effects of public capital cannot be captured in their entirety by the direct effects for each region from public capital installed in the region itself. . . . Ultimately, the aggregate effects are due in almost equal parts to the direct and spillover effects of public capital.

Using evidence on fixed-line telecommunications networks, Röller and Waverman (2001) argue that such networks have a positive causal link on economic growth but typically only when a near universal service is provided. This characteristic they attribute to 'network externalities: the more users, the more value is derived by those users' (*ibid*., p. 911). Vickerman (2007, p. 598) similarly argues that the cost–benefit analysis (CBA) of large-scale infrastructure projects, 'needs to be able to incorporate network impacts which are notoriously difficult to identify and model'.

## III. Micro view of infrastructure delivery

In contrast to the aggregate and network-level preoccupation of macro-studies, a contrasting micro-level strand of literature has developed in parallel in planning and management. The micro-level literature is based on evidence from project-level case studies and larger datasets of actual outcomes of infrastructure mega-projects in terms of cost, time, and benefit performance (Flyvbjerg *et al*., 2002, 2003, 2005, 2009; Flyvbjerg and Budzier, 2011; Ansar *et al.*, 2014).

Pickrell (1992) studied rail transit projects in US cities such as Washington DC's Metro system. He found the 'forecasts that led local officials in eight US cities to advocate rail transit projects over competing, less capital-intensive options grossly overestimated rail transit ridership and underestimated rail construction costs and operating expenses' (p. 158). Pickrell's evidence on the gap between the aspiration and reality of infrastructure projects formed the basis of and lent credence to the notion of 'lying with numbers' (Wachs, 1989).

In a similar vein of thought, Flyvbjerg (1998) undertook a richly detailed case history of the Aalborg Project—a project to redevelop the downtown area of Denmark's third-largest municipality. Flyvbjerg (1998, p. 225) found that, even in a transparent democracy like Denmark's, although the aspiration of the Aalborg project 'was based on rational and democratic argument. During implementation, however, when idea met reality . . . It disintegrated into a large number of disjointed sub-projects, many of which had unintended, unanticipated and undemocratic consequences'. The unfavourable outcomes of the Aalborg project led Flyvbjerg and colleagues (Flyvbjerg *et al*., 2002, 2003, 2005) to publish a series of works exploring the 'anatomy of risk' in infrastructure megaprojects in much larger datasets that, unlike Pickrell's study and for the first time, allowed statistically valid conclusions. Using data from 258 transport infrastructure projects, Flyvbjerg *et al*. (2002, p. 279) found that the cost estimates used to decide whether such projects should be built were 'highly and systematically misleading. Underestimation cannot be explained by error and is best explained by strategic misrepresentation, that is, lying'.

The concept of 'strategic misrepresentation' has its conceptual underpinning in agency theory (Eisenhardt, 1989; Flyvbjerg *et al.*, 2009). Flyvbjerg *et al*. (2002, 2003) reasoned that if inaccurate cost estimates were a consequence of technical causes, errors in overestimating costs would have been of the same size and frequency as errors in underestimating costs. Moreover, in line with models in economics such as 'rational expectations', forecasting errors, if they were technical, would approximate a more or less symmetrical distribution around a stable mean of zero for a large sample of projects. But neither turns out to be the case. Forecasting errors are systematically biased towards adverse cost overruns with a mean significantly different from zero across project types. Similarly, neither the frequency nor the magnitude of cost overruns has improved over the last 70 years (Flyvbjerg *et al.*, 2002). Flyvbjerg *et al.* (2005) found a similar pattern of adverse outcomes in benefit estimates of infrastructure projects—ridership volumes, for example of urban rail projects, systematically fell short of their targets.

Building on evidence from the large datasets and interview data from close dialogue with practitioners in the field of infrastructure delivery, Flyvbjerg (2005, p. 18) proposed that infrastructure megaprojects were the progeny of a 'Machiavellian formula', which paraphrasing goes as follows:

In order to get an infrastructure project built:

(under-estimate costs)
+ (over-estimate revenues)
+ (under-value environmental and social impacts)
+ (over-value wider economic development effects, or spillover effects)
= (win project approval).

The result of these realpolitik tactics in the appraisal, selection, and building of infrastructure projects is an unhealthy 'survival of the unfittest' by which the 'worst infrastructure gets built' (Flyvbjerg, 2009).

The evidence of systematic cost overruns and benefit shortfall has also invited interest from researchers in the field of psychology. Dan Lovallo with Daniel Kahneman wrote in the *Harvard Business Review* (2003, p. 58):

> When forecasting the outcomes of risky projects, executives all too easily fall victim to what psychologists call the planning fallacy. In its grip, managers make decisions based on delusional optimism rather than on a rational weighting of gains, losses, and probabilities. They overestimate benefits and underestimate costs. They spin scenarios of success while overlooking the potential for mistakes and miscalculations. As a result, managers pursue initiatives that are unlikely to come in on budget or on time—or to ever deliver the expected returns.

Flyvbjerg (2003, p. 121), in an invited comment in *Harvard Business Review* on Lovallo and Kahneman (2003), argued,

> Their look at overoptimism, anchoring, competitor neglect, and the outside view in forecasting is highly useful to executives and forecasters. But Lovallo and Kahneman underrate one source of bias in forecasting—the deliberate 'cooking' of forecasts to get ventures started. My colleagues and I call this the Machiavelli factor.

Since that debate, scholars in management and psychology have come to view overoptimism (delusion) and strategic misrepresentation (deception) as complementary rather than alternative explanations of the failure of large infrastructure projects. In practice it is often difficult to disentangle the two explanations. Research has typically focused on situations where the explanatory power of one of the two models is greater. For example, learning or adaptive rationality serves to minimize delusion (Gigerenzer, 2002). Opportunities for learning exist 'when closely similar problems are frequently encountered, especially if the outcomes of decisions are quickly known and provide unequivocal feedback' (Kahneman and Lovallo, 1993, p. 18). Whereas, the problem of strategic deception occurs when incentives are misaligned. The underlying causes of these misalignments are differences in goals, incentives, information, or time horizons between principals and agents (Flyvbjerg *et al.*, 2009, Figure 3).

From a micro perspective, the implication of delusion and deception in infrastructure projects is that these twin forces profoundly undermine the economic return of the individual infrastructure projects that get built. Viewed from the lens of strategic misrepresentation and over-optimism, the Aschauer-style macro-studies appear implausible. We address ourselves to the paradox of how can poor outcomes of infrastructure investment projects—as reported in micro-level studies—amount to economic welfare in the aggregate?

In what follows we study this paradox for China where spectacular growth in nominal GDP has gone in tandem with an unprecedented investment programme (see Figure 1).

## IV. Methods and data

The impact evaluation of project-level infrastructure investments is methodologically a challenging field that is garnering increased attention and creativity (Duflo and Pande, 2007; Estache, 2010; Dinkelman, 2011; McKenzie, 2011; Hansen *et al.*, 2013). Methodological challenges—such as working through a long causal chain, or

**Figure 1:** Gross capital formation (% of GDP) in China versus selected regions

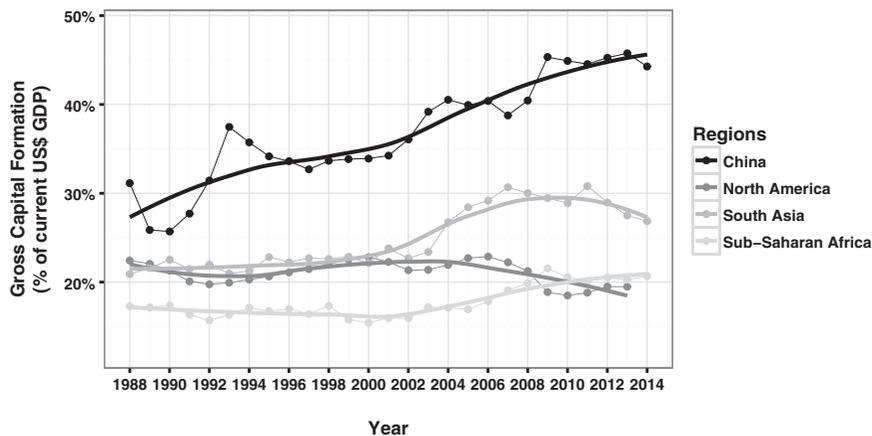

*Source*: World Bank, *World Development Indicators* as of 17/02/2016 update. Indicator name: Gross capital formation (% of GDP); available at http://data.worldbank.org/indicator/NE.GDI.TOTL.ZS

identifying a reasonable counterfactual, or simply identifying a valid and reliable dataset—help explain why bottom-up empirical research on the outcomes of infrastructure projects and their link with macro-level economic performance has been limited in scholarly economics journals.

Our approach has been to collect data on the performance of a large sample of investments to understand whether each of the projects generated economic value, i.e. a benefit-to-cost ratio equal to or greater than one (BCR ≥ 1.0). To this end, we collected data on the actual, *ex post* outcomes related to the benefits, cost, and time of a sample of 95 road[2] and rail[3] infrastructure projects in China built from 1984 to 2008 across 19 (out of 22) provinces, four (out of four) municipalities, and four (out of five) autonomous regions. This is the largest dataset of its kind on China's infrastructure that exists. The portfolio is worth US$52 billion (2010 RMB equivalent) or roughly US$65 billion in 2015 prices. All transport projects for which valid and reliable cost and schedule data could be found were included in the sample. Of the 95 projects, 74 are road and 21 rail projects. Figure 2 presents an overview of the sample.

Even under the best of circumstances, it is difficult to find valid and reliable data on the performance of infrastructure investments (Flyvbjerg *et al*. 2002, 2005; Ansar *et al.* 2014). In China, such difficulties are compounded (Roy *et al.*, 2001; Stening and Zhang, 2007; Quer *et al.*, 2007). Hurst (2010, p. 175) notes:

> Despite all [the] apparent advantages [of large-N quantitative studies in China], several factors detract from the appeal of such methods. . . . First, there is the issue of practicability. It is not easy to obtain good quantitative data in China, just as obtaining good qualitative or interview data is difficult as well. But getting quantitative data is more costly in financial terms . . . and the process of gathering quantitative data is even more tightly controlled for foreign researchers than the gathering of qualitative data. . . . Second, there are some variables in China about which it is exceptionally difficult to obtain or collect accurate quantitative data.

To overcome the challenge of finding reliable data on the outcome of forecasts on important decisions in China, our empirical strategy relied on documentary evidence contained in the loan documents—*ex ante* planning and *ex post* evaluation, or 'retrospective reports' (Miller *et al.*, 1997)—of International Financial Institutions (IFIs), namely (i) the Asian Development Bank (ADB) and (ii) the World Bank, which were the most reliable sources of data we could find on infrastructure projects in China. We discuss the pros of cons of IFI documents in Ansar *et al.* (2012). Our data collection approach also gave us the opportunity to develop more detailed case histories to richly illustrate statistical results and identify causal mechanisms. Like our quantitative data on China's infrastructure, the qualitative case histories were also drawn from documentary evidence.

---

[2] Road projects in China can be typically further subdivided into four sub-categories: (i) four-lane tolled inter-city expressways; (ii) highways, i.e. roads that are classified as class I (25.5m wide), class II (12m wide), or class III (8.5m wide) roads in China; (iii) urban roads and urban road bridges such as the Shanghai's inner ring road or Yangpu bridge; (iv) unclassified rural roads.

[3] We only report data on conventional inter-city heavy rail lines. Although China has built the world's longest high-speed rail network and urban rail networks, we do not yet have data on their outcomes.

**Figure 2:** Sample distribution of 95 transport projects in China (1984–2008), worth US$52 billion (in 2010 RMB equivalent)

**Regional Location within China**

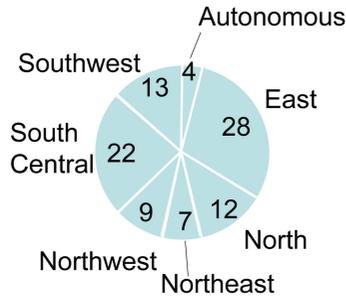

**Project Type**

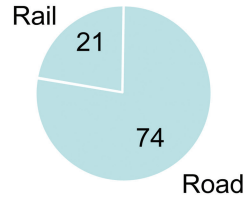

**Financing Agency by Project Type**

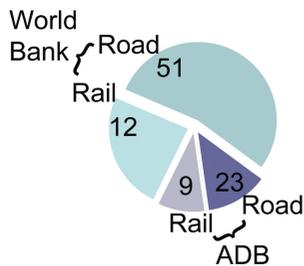

**Project Vintage (By Start Year)**

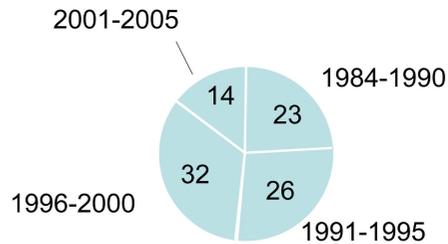

**Actual Project Cost** (USD millions – 2010 RMB equivalent), percent

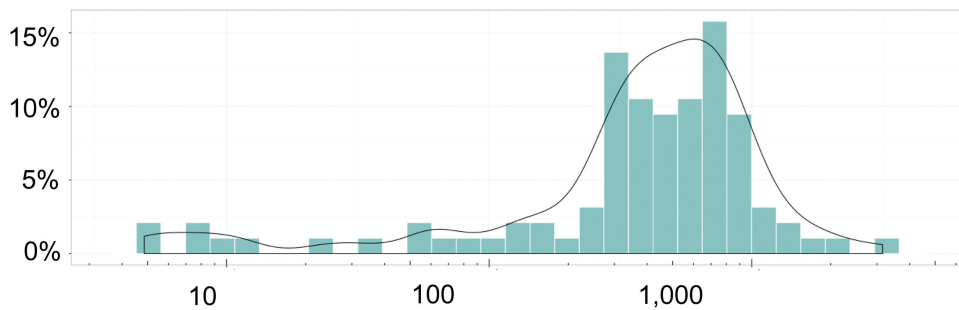

*Source*: Authors' database.

*Measures*

The measures we used are as follows.

   *Cost performance*, cost underrun or overrun, was measured as the actual outturn costs expressed as a ratio of estimated costs. Costs were measured as construction costs comprising the following elements: right-of-way acquisition and resettlement; design engineering and project management services; construction of all civil works; equipment purchases excluding rolling stock. Actual outturn costs are defined as real, accounted construction costs determined at the time of project completion in the local currency—i.e. renminbi (RMB) in China. Estimated costs are defined as budgeted, or forecasted, construction costs in RMB at the time of the decision to build. The year of the date of the decision to build a project is the base year of prices in which all RMB-denominated estimated and actual constant costs have been expressed in real terms—i.e. with the effects of inflation removed. We also exclude from our calculations debt payments, taxes, any *ex post* environmental remedial works. This makes comparison of estimated and actual costs of a specific project a like-for-like comparison.

*Schedule performance* was measured as the ratio of the actual project implementation period to the estimated project implementation period. It therefore measures slippage of the construction programme. The start of the implementation period is taken to be the date of project approval by the main financiers, and the end is the date of full commissioning.

*Benefits performance* was measured as forecasted versus actual traffic. Since many of the 74 roads in our sample were divided into sub-sections, the traffic data was available to us on a road sub-section level. Consequently, for the purposes for benefits performance the number of observations was 156, of which 137 projects were roads and 19 rail projects. Like cost underrun or overrun, benefit excess or shortfall is the actual benefits expressed as a ratio of estimated benefits for each relevant year of operation (see Flyvbjerg *et al.* (2005) for details). We measured road traffic as number of vehicles (usually as 'annual average daily traffic', AADT). Rail traffic was measured as freight tonne-km (millions), thousands of tons, or passenger-mm (millions) whichever was appropriate and available—the choice of measures related to benefits is discussed further in the lead-up to *Policy Proposition 2* below.

*A note on international comparisons*

The cost, schedule, and benefits measures used in our study closely follow those used in the studies of Pickrell (1992), Flyvbjerg *et al*. (2002, 2003, 2005, 2009), and Flyvbjerg (2005, 2009). The present study is part of a larger on-going investigation, conducted by the authors, of large-scale projects around the world. The concerted research effort is yielding growing sample sizes on the outcomes of global infrastructure projects (see, for example, Flyvbjerg and Sunstein (forthcoming) and Cantarelli *et al.* (2012)). This allows our data sample of Chinese transport projects to be directly comparable with data not only in the earlier publications but also the more recent research. For cost and schedule performance, we are thus able to report valid international comparisons between China and many of the rich democracies in the Americas, Europe, and Asia Pacific for the first time.[4] To this end we conducted comparative statistical analysis between our China dataset (95 road and rail projects) with a dataset of projects built in rich democracies—806 road and rail projects (for cost data) and 195 projects (for schedule data). The composition of the combined international dataset is reported in Table 1. We will revisit the findings reported here—particularly the international comparative analyses—in future endeavours as more data become available. Comparative international analysis of benefit performance is a pressing area requiring further research.

---

[4] We use the Organization for Economic Cooperation and Development (OECD) members as proxy for rich democracies: http://www.oecd.org/about/membersandpartners/

**Table 1:** International projects by project type and country

|  | Cost performance data | | |
|---|---|---|---|
|  | Rail | Road | Total |
| **China** | 21 | 74 | 95 |
| **Rich democracies** | 168 | 597 | 806 |
|   Australia | 0 | 2 | 3 |
|   Canada | 6 | 0 | 6 |
|   Denmark | 4 | 21 | 27 |
|   France | 0 | 18 | 19 |
|   Germany | 11 | 1 | 15 |
|   Greece | 0 | 2 | 2 |
|   Ireland | 0 | 27 | 29 |
|   Japan | 6 | 1 | 8 |
|   Korea, S. | 2 | 138 | 140 |
|   Mexico | 1 | 1 | 2 |
|   Netherlands | 29 | 42 | 77 |
|   Norway | 5 | 15 | 31 |
|   Slovenia | 0 | 36 | 36 |
|   Spain | 2 | 0 | 2 |
|   Sweden | 26 | 60 | 86 |
|   Switzerland | 0 | 0 | 10 |
|   UK | 12 | 189 | 202 |
|   USA | 64 | 44 | 111 |
| **Grand total** | **189** | **671** | **901** |
|  | Schedule performance data | | |
|  | Rail | Road | Total |
| **China** | 21 | 74 | 95 |
| **Rich democracies** | 65 | 118 | 195 |
|   Australia | 0 | 11 | 14 |
|   Canada | 0 | 1 | 1 |
|   Denmark | 3 | 12 | 15 |
|   Germany | 5 | 0 | 6 |
|   Ireland | 0 | 3 | 3 |
|   Japan | 4 | 0 | 4 |
|   Korea, S. | 1 | 0 | 1 |
|   Netherlands | 1 | 42 | 49 |
|   Norway | 1 | 1 | 3 |
|   Slovenia | 0 | 36 | 36 |
|   Spain | 2 | 5 | 7 |
|   Sweden | 1 | 0 | 1 |
|   UK | 7 | 3 | 10 |
|   USA | 40 | 4 | 45 |
| **Grand total** | **86** | **192** | **290** |

*Source*: Authors' database.

## V. Results

Here we report results on the cost, schedule, and benefit outcomes for the 95 transport infrastructure projects in China in our reference class. We collected data on the 24 variables listed in Table 2.

**Table 2:** Variables and characteristics of major transport projects in China

**Basic project features**
   Road or rail project (dummy variable)
   New project or upgrade (dummy variable)

**Physical scope and size**
   Length of road or rail (kilometres)
   No. of lanes and or tracks
   Percent of road or rail underground, elevated, and at grade, respectively, totalling 100 per cent

**Cost**
   Estimated project cost (constant millions of RMB in 2010 prices)
   Actual project cost (constant millions of RMB in 2010 prices)
   Cumulative inflation contingency (percentage)

**Time**
   Year of final decision to build
   Estimated implementation schedule (months)
   Year of start of full commercial operation
   Actual implementation schedule (months)

**Benefit**
   Estimated traffic (as freight tonnes for rail and number of vehicles for roads)
   Actual traffic

**Procurement and financing**
   Estimated project foreign exchange costs as a proportion of estimated total project costs (percentage)
   Competitiveness of procurement process, amount under international competitive bidding as a proportion of estimated total project costs (percentage)*
   Main contractor is from China (dummy variable)*
   World Bank or ADB financed project (dummy variable)
   World Bank and/or ADB financing—proportion of estimated project cost (percentage)
   Project received central government subsidy (dummy variable)

**Economic and political context variables**
   Administrative level (central, provincial, prefecture, county, township)
   Name of province in which project nested (where relevant)
   Index of political status of province in China (where relevant)
   GDP of China (current US dollars)
   *Per capita* income of China in year of project approval (2000 constant US dollars)
   Average actual cost growth rate in China over the implementation period—the GDP deflator (percentage)
   Manufacturers Unit Value index of actual average cost growth rate for imported project components between year of loan approval and year of project completion
   Three-year moving average of the inflation rate in China (percentage)
   Actual average exchange rate depreciation or appreciation between year of formal-decision-to-build and year of full commercial operation (percentage)

*Note*: * Denotes variables with a large number of missing values
*Source*: Authors' database.

*Cost performance*

With respect to cost performance, we make the following observations:

- 75 per cent of transport projects suffered a cost overrun in constant local currency terms.
- Actual costs were on average 30.6 per cent higher than estimated costs, with a median of 18.5 per cent indicating that the distribution of costs had a heavy skew to the right (i.e. going over budget). A Wilcoxon signed rank test of overall cost neutrality provided conclusive evidence that costs were systematically biased towards

   underestimation (p < 0.0001); the median cost underestimation was 27.6 per cent of budget against 9.01 per cent for cost overestimation—a difference that is overwhelmingly significant (p < 0.0001). There is a heavy bias towards adverse outcomes as shown in Figure 3.
– Wilcoxon signed rank tests suggested that both road and rail projects suffered cost overruns significantly above zero in China reported in Table 3. Roads, however, performed better with a lower average and median cost overrun than rail projects as summarized in Table 3 (see also Figure 3). Similarly, seven out of 10 roads went over budget, whereas nine out of 10 rail projects went over budget.
– Using a non-parametric Wilcoxon test, we tested whether the distribution of cost overruns in transport projects in China ($N = 95$) are different from rich democracies ($N = 806$; see the methods section). We found no significant differences in cost overruns between China and rich democracies ($W = 38712$, $p = 0.8591$)—i.e. based on our sample China's cost performance is no better or worse than that of rich democracies. This result puts in doubt the oft-repeated hypothesis among scholars and the media that autocratic political systems, such as China's, may have an edge in controlling the delivery of infrastructure. This intriguing line of enquiry demands further research, particularly because we could not control for covariates such as project size or year of project start. This result, therefore, should be treated as a preliminary outcome requiring further research.

*Schedule performance*

With respect to schedule performance, we make the following observations.

   – On average road and rail projects in China took 4.3 years to build.
   – Roads took 3.9 years on average from start to completion. Rail projects took longer to implement with an average implementation schedule of 5.5 years. Note that these schedules did not take into the account the lead times in preparing projects.

**Figure 3:** Density trace of costs overruns in constant RMB by project type and mean (vertical lines)—road ($n = 74$) and rail ($n = 21$)

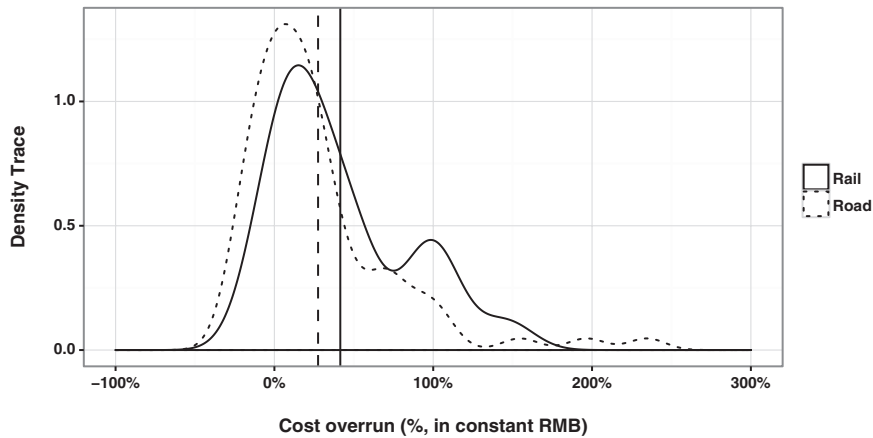

*Source*: Authors' database.

- Breaking the schedule delay by project type revealed that rail projects suffered a 25 per cent ($SD = 36.7$) schedule overrun, which is statistically significantly biased above zero (non-parametric test, $p = 0.0053$). Roads, however, do not incur a schedule delay on average ($SD = 29.5$).
- Figure 4 shows the average schedule overrun (%) by project type in China.
- In terms of actual construction time, Chinese projects have a shorter actual duration from the decision to build to completion than those of rich democracies. On average Chinese projects took 4.3 years (median = 4.0) and projects in rich democracies took on average 6.9 years (median = 6.0). This difference is statistically significant (non-parametric Wilcoxon test, $W = 28365$, $p < 0.001$).
- Similarly, in terms of schedule overrun China performed better than rich democracies ($W = 12087$, $p < 0.001$). The average schedule overrun in rich democracies was +42.7 per cent (median = +23.0 per cent) compared to Chinese projects' average of +5.9 per cent (median = 0.0 per cent). Only one in every two projects encountered a schedule delay in China compared to seven out of 10 in rich democracies.

This finding can be explained from two competing perspectives. Psychological theories might suggest that Chinese planners are less optimistic about the time it takes to get a project done than their counterparts in democratic countries. However, if Chinese

**Table 3:** China—cost overruns by project type

| Project type | Number of cases (N) | Average cost overrun (%) | Standard deviation | Level of significance ($p$) | Median cost overrun (%) | Frequency of projects cost overrun (%) |
|---|---|---|---|---|---|---|
| Road | 74 | 27.5 | 47.7 | < 0.0001 | 16.1 | 70 |
| Rail | 21 | 41.5 | 43.2 | < 0.0001 | 28.5 | 90 |
| Total | 95 | 30.6 | 46.9 | < 0.0001 | 18.5 | 75 |

*Source*: Authors' database.

**Figure 4:** Density trace of schedule overruns by project type and mean (vertical lines)—road ($n = 74$) and rail ($n = 21$)

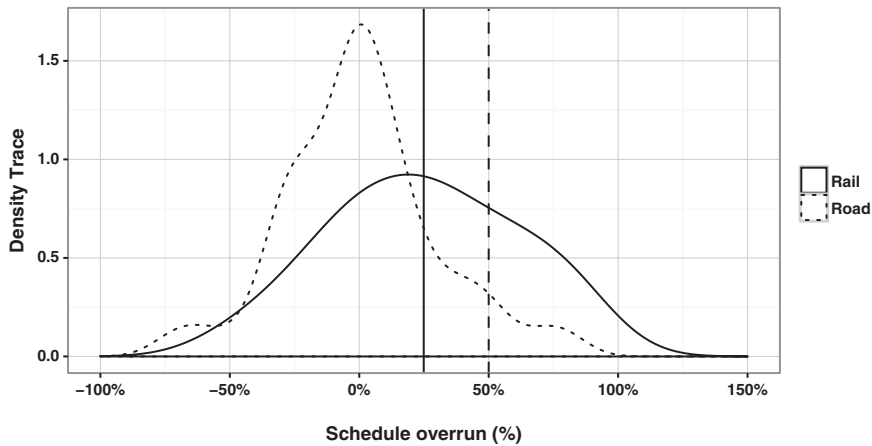

*Source*: Authors' database.

planners were less optimistic in general, then one would expect better cost forecasting performance as well, which is not the case. In contrast, agency theories might suggest that incentives for Chinese cadres and contractors are such that building as quickly as possible is rewarded even if performance in other areas such as cost, quality, safety, environmental impact, or public consultation processes is allowed to slack.

There is greater qualitative support for the agency theory's perspective with respect to this finding. For example, China's approach to land acquisition and population resettlement is heavy-handed (Ren, forthcoming). Similarly, quality and safety issues in China's infrastructure projects are not uncommon (Zou *et al.*, 2007). For instance, road fatalities in China are some of the highest in the world—18.8 fatalities per 100,000 inhabitants per year, compared to 2.9 in the UK, according to the World Health Organization (WHO, 2015, Table A2)—due in large part to poor technical design and road quality issues (Ameratunga *et al.*, 2006; World Bank, 2008).

With growing importance of issues such as cost, safety, or the environment in China, Chen (2014) argues that construction schedules and delays will also increase:

> delays used to be greatly frowned upon when it came to major building projects in China. . . . A 1,300km high-speed rail line between Beijing and Shanghai that was due to take five years was finished in just over half the time.

But now projects such as China's first passenger jet, the C919, or nuclear reactors are running late. Chen (2014) writes,

> Economics professor Zhao Jian, from Beijing Jiaotong University calls it the 'new normal'—a term used by President Xi Jinping . . . [because] of less pressure from the top to finish projects as fast as possible. . . . Han Kecen, of the Shanghai Airplane Design and Research Institute and administrative commander of the [delayed] C919 airliner project [said] 'Time is not the most important element; the top priority is to guarantee the safety of the plane'. . . . Words that would not have been heard a decade ago, when the 'old normal' in China was speed, first and foremost.

In brief, China has built infrastructure at impressive speed in the past but, it appears, by trading-off due consideration for quality, safety, social equity, and the environment. The frequent laments by politicians in rich democracies that public consultation processes amount to 'dithering' are misguided (see Johnson, 2013).

*Benefits performance*

With respect to benefits performance, we were able to gather traffic data for 156 projects in China of which 137 projects were roads and 19 rail projects.

Prima facie, Chinese projects did not have significant traffic shortfalls, with an average shortfall for road and rail combined that was only –5.0 per cent with a large standard deviation ($SD = 61.4$). However, the –5.0 per cent is an average of two extremes that are both undesirable. Thus a majority of routes witnessed paltry traffic volumes while a few routes were highly congested, as follows.

– Approximately, two-thirds (64.7 per cent) of the 156 projects had benefit shortfalls, i.e. actual traffic volumes were biased below forecasts. The magnitude of bias in these instances is staggering: the average traffic shortfall for these routes was –41.2 per cent ($SD = 23.1$). Some routes received less than 20 per cent of their forecast traffic.

- In contrast, the remaining third of the projects (35.3 per cent) enjoyed too much of a good thing, experiencing an average traffic surplus of +61.4 per cent ($SD = 53.7$) causing congestion problems. Indication of congestion—traffic volumes up to three times their original forecasts—is found for urban roads and bridges and expressway routes that connected very large and nearby cities such as the Beijing–Tianjin–Tanggu Expressway.

Both extremes, which can be seen in Figure 5, are equally undesirable, because large unused capacity equals waste, as does too little capacity, seeing as it is considerably more expensive to add capacity to existing fully used routes than it is to build the capacity up front.

We acknowledge that the volume of traffic, as measured above, is a proxy for benefits. In order to calculate the net benefits more fully, one not only requires volumetric traffic data but also toll price data. Toll data at the project level are, however, not easily available—a gap that future research may look to fill. Nevertheless, we dealt with this potential gap by closely analysing the toll price structure of a subset of the 95 cases we studied. Of the 74 road projects we studied, 85 per cent were toll roads. We shortly turn to the case of the Yuanjiang–Mohei 'YuanMo' expressway that shows that actual prices ($P$) achieved for tolls were also biased below their estimate at appraisal along with the traffic volumes ($Q$). We had detailed before and after financial and economic evaluation models available to us for the YuanMo expressway. Hence we could perform more thorough benefit-to-cost ratio (BCR) calculations for the YuanMo case than was possible for other projects. However, by using pecuniary cost overruns (or underruns) data but volumetric benefits data we were able to estimate the *ex post* BCRs of other projects in our sample. We are thus able to form reasonable judgements on the economic viability of individual projects.

**Figure 5:** Density trace of benefit shortfall or excess (%) by project type and mean (vertical lines)—road ($n$ = 137) and rail ($n$ = 19)

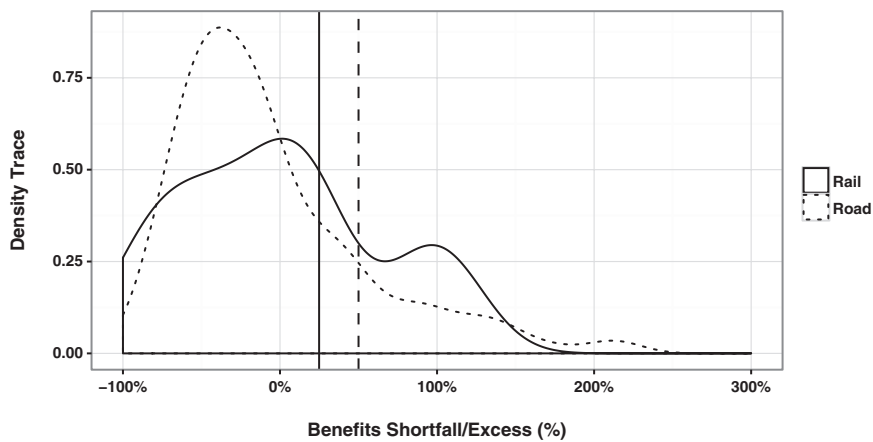

*Source*: Authors' database.

## VI. Summing up the evidence

Putting the evidence of cost overrun and benefit shortfalls together paints a grim picture of outcomes of large transport projects in China. The infrastructure investments we studied were based on *ex ante* cost–benefit analysis. The final decision to go ahead with a project investment was based on the belief that the BCR would exceed 1.0—i.e. the investment in the project would produce a positive net present value (NPV) and hence be economically viable.

In the reports we studied for China, the typical BCR for transport projects was 1.4 to 1.5, which is broadly in line with many other physical infrastructure assets such as large dams, road, rail, bridge, or tunnel capital investments (Fan and Chan-Kang, 2005, p. 44; Ansar *et al.*, 2014; NAO, 2014). In other words, planners expected the net present benefits to exceed the net present costs by about 40–50 per cent.

In the case of the YuanMo expressway, for example, the *financial* BCR—based on the present value of total accounting outlays and toll revenues—was 1.10 at a 10 per cent discount rate (ADB, 1999, p. 82, Table A17.1). The project's financial internal rate of return (FIRR) was 10.9 per cent, a small margin above the hurdle rate of 10 per cent. In order to justify the project, the ADB (1999) business case also incorporated wide economic benefits, which pushed the expected *economic* BCR to 1.5 (see ADB, 1999, p. 87, Table A18.2).

The combined effect of benefit shortfalls and cost overruns, however, pushed the BCR (not only the narrower financial but also the wider economic measure) below 1.0. For YuanMo, incomplete technical design at appraisal, difficulties with the topography and the geology, and land acquisition and resettlement issues pushed the CAPEX costs up 24 per cent (year-of-expenditure RMB). To make matters worse, actual traffic volumes were '49% lower than that forecast at appraisal' (ADB, 2006, p. 43). Even after an 8-year ramp-up period, as of May 2011—the latest date for which the traffic count was available to us—the first year traffic forecast was yet to be met.

The lost revenues due to lower traffic volumes for YuanMo expressway were exacerbated by a 53 per cent shortfall in the forecasted toll rates. 'Giving due consideration to affordability by road users and other social and economic impacts, the existing weighted, average toll rates [were] lower than those proposed at appraisal,' explained the ADB (2006, p. 11). A 49 per cent shortfall in traffic volumes and a 53 per cent shortfall in toll prices combined to yield revenues that were approximately a quarter of the original forecasts.

Taking into account the cost overrun and revenue shortfall, we recalculated the BCR of the YuanMo project. The financial BCR, according to our revised estimates, fell to 0.2. The economic BCR, even after including generous provisions for wider benefits, only improved to 0.3. Both calculations are based on the assumption that the traffic volumes and toll revenues would not dramatically improve. The available data on the YuanMo project's performance over the last 12 years support these assumptions. There was no plausible scenario in which—after suffering a 24 per cent cost overrun, a 49 per cent traffic shortfall, and a 53 per cent toll-price shortfall—the YuanMo expressway could yield a positive return. The project destroyed economic value.

It was possible to map both the cost and benefit data for 65 projects (from our sample of 95 observations). Like the YuanMo case, 55 per cent of the projects had an *ex post* BCR lower than 1.0—i.e. these projects were economically unviable at the outset of their operational lives as shown in the lowest region in Figure 6. A majority of these value-destroying projects suffered the double whammy of a cost overrun and a benefit shortfall. Another 17 per cent of the projects generated a lower-than-forecasted BCR shown in the middle region of Figure 6. Any future risks, such as greater-than-expected operation and maintenance costs, can impair the future economic viability of these projects. Finally, only 28 per cent—i.e. less than a third of our sample—can be considered genuinely economically productive.

Generalizing from our sample, evidence suggests that over half the infrastructure investments in China made in the last three decades have been NPV negative. Far from being an engine of economic growth, a typical infrastructure investment has destroyed economic value in China due to poor management of risks that impact cost, time, and benefits.[5] We advance:

*Hypothesis 1. Due to a propensity to cost overruns and benefit shortfalls, the typical infrastructure investment destroys economic value.*

*Policy Proposition 1. Less is more. Policy-makers should only commit scarce public resources to infrastructure alternatives that, even after accounting for potential cost overruns and benefit shortfalls, produce positive economic value.*

Proponents of infrastructure investments often argue that, even if individual projects such as the YuanMo expressway yield negative NPVs, the benefits of a network will outweigh the cost of building the network. Although an appealing argument, this is unlikely to hold in the real world. First, the business cases of individual infrastructure projects are justified on the basis of their NPV being positive. When the NPV becomes negative in reality, planners go to some length to obfuscate the inconvenient truth—a persistent and insidious feature of infrastructure investments (Wachs, 1989). Second, construction cost or time and traffic volumes are tangible and quantifiable indicators. Given the systematic biases in these simple metrics, more complex

**Figure 6:** Proportions of projects by *ex post* estimates of BCRs (*n* = 65)

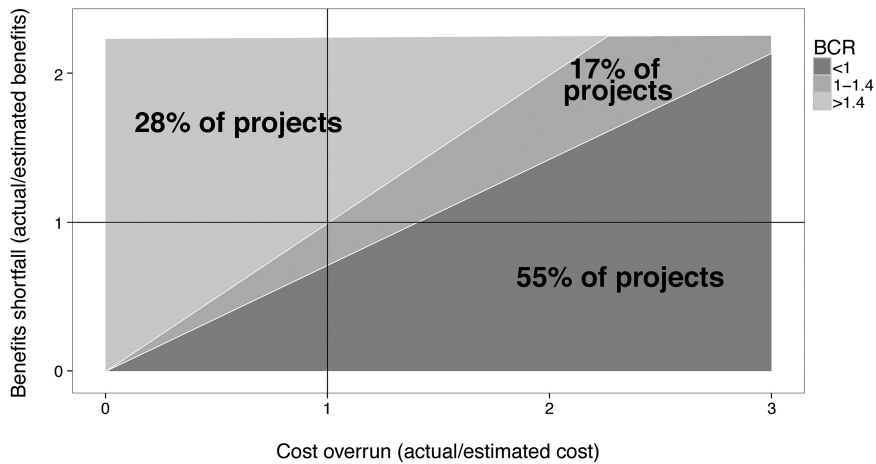

*Source*: Authors' database.

---

[5] Only six out of 66 projects can be considered outright successes where benefits greatly exceeded costs—this suggests a composite success rate of less than 10 per cent. Venture capital investors, not governments, are meant to take on endeavours with such risky pay-offs.

metrics, such as network and spillover effects, will be prone to a greater degree of delusion and deception.

How ought policy-makers then account for the benefits of infrastructure projects? In the case of transport infrastructure, benefits of projects are typically enumerated across many dimensions. Promoters claim that a new project will create new jobs, or cause the value of land adjacent to a project to appreciate, or provide value of time savings for potential end-users. Our broader evidence from China and the deeper case studies, of which the YuanMo expressway is an example, suggest that the wider the net of benefits policy-makers attempt to cast, the weaker the business case of the proposed infrastructure. Benefits, such as value of time savings or increased land values, do not come about unless the forecast traffic volumes materialize. Actual traffic is thus the most concrete and fool-proof gauge of the actual benefits of a transport project. If the basic traffic does not materialize, the rest of the benefits are also unlikely to emerge.

Wider benefits are a poor guide to infrastructure investment decision-making. Vickerman (forthcoming, pp. 22–3) concludes that, (i) wide benefits where they exist, typically account for 10–20 per cent, in addition to direct benefits, (ii) often wider impacts do not exist or are negative, and (iii) where wider positive impacts exist in some regions they could be offset by negative impacts in other regions, reducing the aggregate effect.

In formal terms:

*Hypothesis 2. Direct benefits, e.g. financial cash flows, at the project-level will be a more robust measure of the actual benefits of infrastructure investments than wider economic benefits or network effects.*

*Policy Proposition 2: Instead of enumerating many, potentially obscure, dimensions of future benefits, policy-makers should focus on one simple metric—such as the actual Annual Average Daily Traffic (AADT) or revenues—for infrastructure investments.*

Does China's high-octane investment programme in infrastructure explain its high economic growth rate? The conventional wisdom in economics has tended to present the seemingly obvious answer, which Röller and Waverman (2001, p. 909), using telecommunication networks as an example, neatly summarize: 'investing in telecommunications infrastructure does itself lead to growth because its products—cable, switches, and so forth—lead to increases in the demand for the goods and services used in their production.'

In contrast, the implication of our research is that economists have tended to overstress the need for infrastructure in the economy by dwelling on the link between infrastructure investment and short-term economic growth. It is a given that increased physical capital accumulation (irrespective of whether the investment has a positive or negative NPV) will increase the GDP in the short run as a natural accounting consequence of piling investments (productive or not) into fixed capital. In fuelling economic growth today by excessive capital accumulation, policy-makers risk suffocating the possibility of steadier and more resilient future economic growth that comes from greater efficiency and productivity of using scarce factors of production.

Banister and Berechman, (2000, pp. 149–50) corroborate our observation: 'The nature of the causality between transport infrastructure development with economic growth is rather equivocal with respect to direction, functional relationships, and effect of intervening variables'. With respect to China, Huang and Khanna (2003) and Huang (2006, 2008) also stress the direction of this causality. Huang (2006) argues:

> This is [a] 'China myth'—that the country grew thanks largely to its heavy investment in infrastructure. This is a fundamentally flawed reading of its growth story. In the 1980s, China had poor infrastructure but turned in a superb economic performance. China built its infrastructure after—rather than before—many years of economic growth and accumulation of financial resources. The 'China miracle' happened not because it had glittering skyscrapers and modern highways but because bold economic liberalization and institutional reforms—especially agricultural reforms in the early 1980s—created competition and nurtured private entrepreneurship.

China's case carries generalizable policy lessons. A massive infrastructure investment programme is not a viable development strategy in other developing countries such as Pakistan, Nigeria, or Brazil. Policy-makers should place their attention on software and orgware issues (deep institutional reforms) and exercise far greater caution in diverting scare resources to new hardware (physical infrastructure).

## VII. The consequences of profligacy

What are the macroeconomic consequences of systematic cost overruns and benefit shortfalls in infrastructure investments? The macro-view *à la* Aschauer, and particularly the neo-Keynesian school of thought, sees the evidence we have presented about cost overruns and benefit shortfalls in a benign light: paying more for a road only increases the multiplier effect of the investment. In contrast to the macro-view, we will now show that China's investment boom *and* poor project-level outcomes have created pernicious macroeconomic consequences. The most tangible consequences of poor investment decisions have been an accumulation of a destabilizing pile of debt in the economy; unprecedented monetary expansion—even larger than the quantitative easing programmes of the US, the Euro area, the UK, and Japan combined—and subsequent economic fragility to financial crises. Less tangible, but perhaps even more damaging, are the opportunities forgone to build the right infrastructure.

We found a link between China's economic fragility and poor infrastructure project outcomes by considering the following pieces of evidence: (i) the trend in China's gross fixed capital formation; (ii) the trend in debt growth in China and its association with elevated investment levels and cost overruns; (iii) evidence on China's monetary expansion; and (iv) literature on public debt and economic fragility. We now consider each in turn.

China is now the world's biggest spender on fixed assets in absolute terms. Figure 7 presents the gross fixed capital formation (current US$) in China from 1982 to 2014 compared with that in the US, Japan, and Germany—China's three closest rivals in term of annual investment. The scale and speed of China's investment boom are staggering. China spent US$4.6 trillion in 2014 accounting for 24.8 per cent of worldwide total investments and more than double the entire GDP of India. By way of comparison, China's total domestic investment was merely 2.1 per cent of the world total in 1982. The effect of China's economic stimulus programme that started in 2008 is also visible in the data in Figure 7. Undoubtedly, China has been in the grips of the 'biggest investment boom in history' for over 15 years (Flyvbjerg *et al.*, 2009, p. 170).

**Figure 7:** China's investment boom

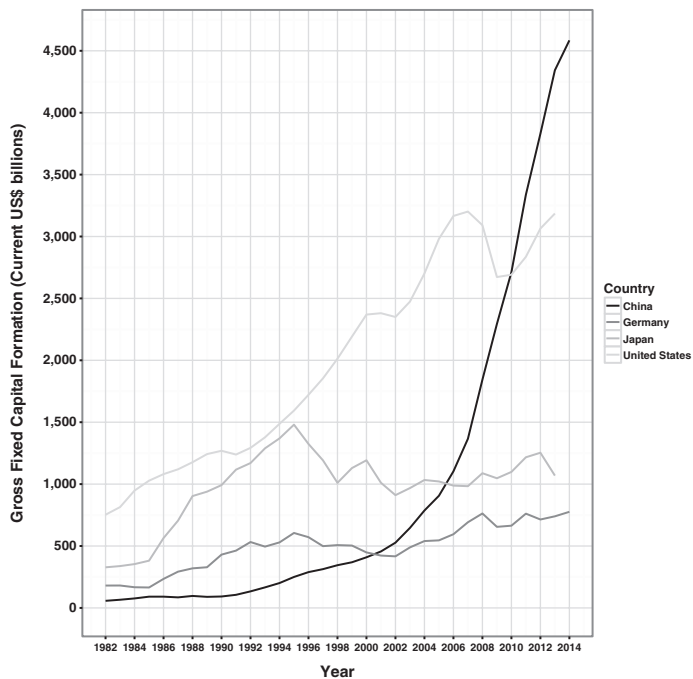

*Source*: World Bank, *World Development Indicators* as of 17 February 2016 update.
Indicator name: gross fixed capital formation (current US$). http://data.worldbank.org/indicator/NE.GDI.FTOT.CD

China's investment boom has coincided with a rapid build-up of debt. According to McKinsey (2015), between 2000 and 2014 China's total debt grew from US$2.1 trillion to US$28.2 trillion, in current prices—an increase of US$26.1 trillion, greater than the GDP of the US, Japan, and Germany combined. The growth in China's absolute debt is neck and neck with the total capital investment, which between 2000 and 2014 was cumulatively US$29.1 trillion. The majority of the investments China has made since 2000 have been debt-fuelled (Barnett and Brooks, 2006; Trivedi, 2015).

All segments of the economy—government, corporate, households, and financial—have contributed to the rapid growth of debt in China, as illustrated in the 2000–14 data in Figure 8. However, the biggest increase has come from the financial sector, dominated by the big four state-owned banks in China, whose debt as a ratio of GDP has grown from 7 per cent in 2000 to now 65 per cent of the GDP (McKinsey, 2015). A high proportion of non-performing loans (NPLs) within this growing financial debt pile is a particular worry (Shih, 2004; Li and Ng, 2013).

China's debt has grown at such a fast pace since 2000 that its debt-to-GDP of 282 per cent now exceeds that of many highly indebted advanced economies, e.g. the United States (269 per cent) and Germany (258 per cent) (McKinsey, 2015). China has also become the most indebted of 25 emerging markets, such as Brazil (160 per cent), India (135 per cent), Russia (88 per cent), or Nigeria (46 per cent)—see McKinsey (2015, p. 106).[6]

---

[6] Includes debt of government, non-financial corporations, households, and the financial sector. Data as of Q2 2014 for advanced economies and China; 2013 data for other developing economies.

**Figure 8:** China's growing debt pile (debt-to-GDP, %)

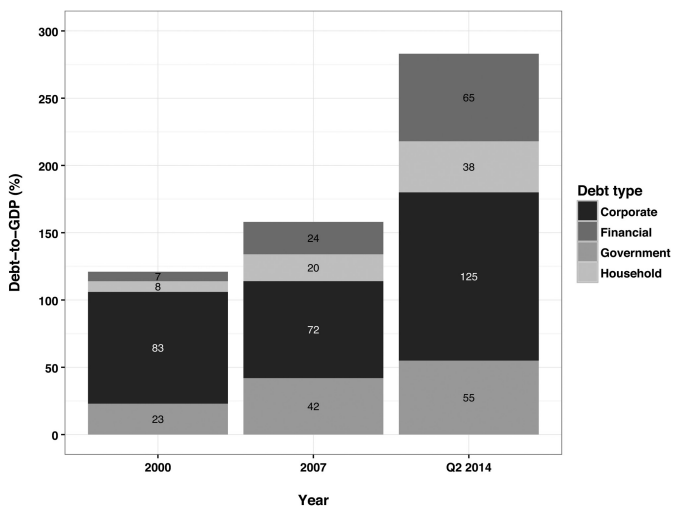

*Source*: McKinsey (2015).

Literature in macroeconomics has tended to focus narrowly on the government portion of total debt (Reinhart *et al.*, 2012; IMF *World Economic Outlook* database). However, government debt is only one of four constituents of the true picture of a country's total indebtedness, which is what really matters. The debt of corporations, households, and financial institutions also has a bearing on the economic prospects of a country (Dynan *et al.*, 2012; Sutherland *et al.*, 2012). By the traditional measure of government debt, China's liabilities seem, at 55 per cent, low when compared to those of Greece (183 per cent) or Japan (234 per cent). However, China's relatively low official government debt-to-GDP ratio understates the actual debt burden the government carries for the following two reasons: corporate borrowing (125 per cent of China's GDP) and financial institution borrowing (65 per cent of China's GDP) are dominated by state-owned or state-controlled entities. The liabilities of these state-linked entities are ultimately governmental in nature if a liquidity crisis were to manifest. Thus:

– China's state-owned enterprises (SOEs)—many of which have been involved in building infrastructure—are the dominant corporate borrowers and continue to receive preferential access to debt (Allen *et al.*, 2005; Chen *et al.*, 2011; Yueh, 2011; Bailey *et al.*, 2012; Cull *et al.*, 2015; Deng *et al.*, 2015). Although SOEs' share of the economy has steadily declined since 1978, even by conservative estimates, state-owned or -linked corporate entities still account for 60–80 per cent of the total outstanding debt (Lardy, 2014, pp. 99–112; Lardy, 2015; cf. Deng *et al.*, 2015). As a low estimate (i.e. 60 per cent) of the US$12.5 trillion of corporate debt estimated by McKinsey (2015, p. 75), an additional US$7.5 trillion can be considered implicitly governmental in nature.
– China's financial institutions are also primarily state-owned. All but a handful of China's banks are directly state-controlled—either by the central government, or through various local and municipal governments and cooperatives, or by majority stakes in joint-stock commercial banks—and estimated to account for over 85 per cent of total bank assets (Boyreau-Debray and Wei, 2004; Firth *et al.*, 2008; 2009; Lardy, 2014, p. 20; Deng *et al.*, 2015, Table III, p. 68). State-controlled banks aside, special purpose local government financing vehicles, trusts, and other shadow financial institutions are also government linked (Chang *et al.*, 2013; Levinger, 2015). Even by conservative estimates, over 90 per cent of China's US$6.5 trillion of financial institutions' debt estimated by McKinsey (2015, p. 83) is implicitly governmental in nature.

Incorporating state-linked corporate and financial institutions' debt, as outlined above, yields estimates of China's true government debt in the range of 190–220 per cent of GDP. As a percentage of GDP, China's is the second-most indebted government in the world—second only to Japan.

The exacerbating role of cost overruns and revenue shortfalls of infrastructure investments is salient in the startling increase of government debt in China. As discussed earlier in the results section, infrastructure investments in China suffered an average cost overrun of 30.6 per cent for our sample of 95 projects. Our reference class—although a small proportion of China's overall investment—carries some basis for extrapolation. Road and rail projects have a lower mean cost overrun than other asset types such as large dams ($M$ = 96 per cent; Ansar *et al.*, 2014) or nuclear power plants ($M$ = 207 per cent; Schlissel and Biewald, 2008, p.8)—investments China has made in abundance. Moreover, World Bank and ADB-financed projects are likely to have better performance than the average Chinese project, making our estimate from the reference class conservative. Finally, we are not incorporating the effects of benefit shortfalls or accrued financing costs. The actual requirement for additional financing is in all likelihood wider than we propose. Given China's cumulative 2000–14 capital investment of US$29.1 trillion, we thus conservatively estimate China's cumulative absolute level of cost overrun at US$8.9 trillion—a vast pool of excess liabilities. China state-owned banks have (willingly and under political pressure) absorbed such liabilities and accepted to roll them forward (Shih, 2004; Chen, 2006; Landry, 2012; Li and Ng, 2013; Sender, 2015). The consequence is 'financial fragility' of the whole system with increased risk of future blowouts (Mankiw, 1986; Davis, 1995; Klemkosky, 2013).

The central bank's willingness to provide additional funding to state-owned banks to prevent default has required an unprecedented monetary expansion. Mandeng (2014) found that China's broad money supply between 2007 and 2013—a period of extraordinary 'quantitative easing' globally—was greater than the rest of the world combined. According to data supplied by Mandeng (personal communication, 24 November 2015), China's M2 broad money grew by US$12.9 trillion in 2007–13—the scale of the increase overwhelms that of the US, where broad money grew by US$3.52 trillion as shown in Figure 9.

The consequences of this build-up of debt and monetary expansion in China, like elsewhere, are not benign. Rapid debt accumulation is positively associated with financial crises (Primo Braga and Vincelette, 2010; Mian and Sufi, 2014). Non-linear negative macroeconomic impacts, such as volatile movements in interest, exchange, and inflation rates; unpredictable movements in asset prices, such as house prices and listed public equities; adverse growth outcomes; rising unemployment from deleveraging; and lack of capital to finance productive investments (Meade, 1958; Checherita and Rother, 2010; Rogoff and Reinhart, 2010; Reinhart *et al.*, 2012; Bowdler and Esteves, 2013): several of these negative consequences are already materializing in China (Financial

Times, 2016; *The Economist*, 2016).[7] China's total *per capita* debt now stands at above US$20,000. The multiple of *per capita* debt to *per capita* annual income for China is 11.5 far greater than that of the US (7.5) or Brazil (8.1), and in line with that of Greece (11.8).[8] Since 2009 Greece has been in the midst of one of the worst sovereign debt crises of recent memory that has hit poorer households particularly hard (Lane, 2012; Matsaganis and Leventi, 2013). Other countries with similarly large *per capita* debt-to-income multiples, such as Spain or Japan, are undergoing debt-induced economic stagnation (Lo and Rogoff, 2014; Baldwin and Teulings, 2014). China has an ageing population (Eggleston *et al.*, 2013). However, Chinese households have a relatively low level of wealth—for example, the average wealth of an adult in the United States is 15.7 times that of an adult in China (Credit Suisse, 2015, pp. 19–22; Xie and Jin, 2015). China's social safety net and pensions system are far less developed than in advanced economies (Eggleston *et al.*, 2013). The ability of the Chinese state to absorb future liabilities of an ageing population—pensions, elderly care, and health—is a pressing concern, and it will depend on how China otherwise spends its money, including on infrastructure.

Public profligacy need not always lead to a dramatic financial crisis (Lardy, 2015). However, Reinhart and Sbrancia (2015) document various measures that governments

**Figure 9:** World broad money creation, US$ trillions, 2007–13

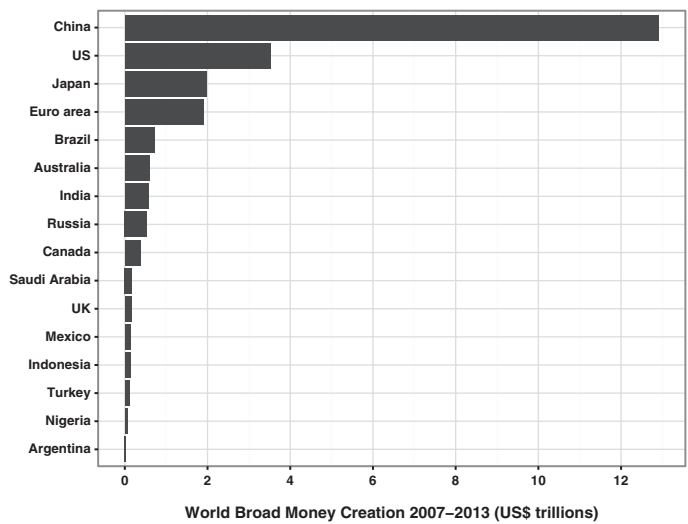

**World Broad Money Creation 2007–2013 (US$ trillions)**

*Source*: Mandeng (2014).

---

[7] On 30 January 2016, in an article entitled 'Grossly Deceptive Plans', *The Economist* wrote: 'On January 19th [2016] China declared that its gross domestic product had grown by 6.9 per cent in 2015, accounting for inflation—the slowest rate in a quarter of a century. It was neatly within the government's target of "around 7 per cent", but many economists wondered whether the figure was accurate.'

[8] *Per capita* annual income for all the countries based on micro-level data on actual household and individual incomes (rather than averages extrapolated from macroeconomic figures) from 131 countries. As of December 2013, median *per capita* annual income of a Chinese adult was US$1,786 compared to US$15,480 for an American or US$2,920 worldwide. See Gallup http://www.gallup.com/poll/166211/worldwide-median-household-income-000.aspx

in a predicament similar to China's have to resort to to 'liquidate debt'. Specifically, insidious forms of 'financial repression' (such as ceilings on interest rates; capital controls; or forcing pension funds to own domestic debt that earns a lower return than a globally diversified portfolio) have to be put in place over long periods of time. For instance, 'China is ratcheting up *ad hoc* capital controls to stem accelerating capital outflows,' reported the *Financial Times* (8 January 2016). Although financial repression can help avert a 'financial meltdown', the measures negatively impact economic welfare (Reinhart and Sbrancia, 2015). New-Keynesian arguments that see public debt in a benign light (Krugman, 2012) are misguided at the level of debt we see in China.

Theory in capital budgeting proposes a simple heuristic: the value of a firm is the sum of the present value of projects in place and the NPV of prospective projects. In particular, 'Capital budgeting processes link CAPX closely to expectations about future firm profits and liquidity' (Souder and Bromiley, 2012, p. 554). Although undoubtedly simplistic, the implication of this rule-of-thumb is salient in the real world. A firm that systematically makes negative NPV investments will run itself aground. This essential capital budgeting framework can also be applied at the national level. A nation's wealth is the sum of the value of all the investments it undertakes. Sacrificing national wealth to build negative NPV investments, such as many of China's infrastructure projects, is economically not prudent.

We hypothesize:

*Hypothesis 3. Cost overruns and revenue shortfalls from poor infrastructure investments will cause a build-up of debt and the risk of economic fragility.*

*Policy Proposition 3. Policy-makers should place a razor-sharp focus on direct project-level cash flows. Choose infrastructure investment alternatives that can generate a positive financial NPV after incorporating the risk of going over budget, time, or under benefits.*

## VIII. Summary and conclusions

The question of whether infrastructure investment leads to economic growth must be answered in the negative. Owing to uncertainty surrounding costs, time, and benefits parameters, a typical infrastructure project fails to deliver a positive risk-adjusted return. There is a common tendency for the benefit-to-cost ratio of major infrastructure investments to fall below 1.0. Such unproductive projects detract from economic prosperity. We thus reject the orthodox theory that heavy investment in infrastructure causes growth.

There is an even more detrimental boomerang effect of overinvestment in infrastructure. Unproductive projects carry unintended pernicious macroeconomic consequences: sovereign debt overhang; unprecedented monetary expansion; and economic fragility.

The primary findings from our datasets are as follows.

– In line with global trends, in China actual infrastructure construction costs are on average 30.6 per cent higher than estimated costs, in real terms, measured from the final business case. The evidence is overwhelming that costs are systematically biased towards underestimation.

- In terms of absolute construction time and schedule overrun China performs better than rich democracies. In democracies politicians seem to have an incentive to over-promise and then under-deliver. China has built infrastructure at impressive speed in the past but, it appears, by trading off due consideration for quality, safety, social equity, and the environment.
- With respect to traffic performance, demand in China represents two extremes. A majority of the routes witness paltry traffic volumes but a few routes are highly congested. Too little and too much traffic of this magnitude both indicate significant misallocation of resources.

The pattern of cost overruns and benefit shortfalls in China's infrastructure investments is linked with China's growing debt problem. We estimate that cost overruns have equalled approximately one-third of China's US$28.2 trillion debt pile. China's debt-to-GDP ratio now stands at 282 per cent, exceeding that of many advanced economies, such as the United States, and all developing economies for which data were available, such as Brazil, India, and Nigeria. Because many corporations and financial institutions in China are state-owned, our revised calculation of China's implicit government debt as a proportion of GDP suggests that China's is the second-most indebted government in the world. Extraordinary monetary expansion has accompanied China's piling debts: China's M2 broad money grew by US$12.9 trillion in 2007–13, greater than the rest of the world combined. The result is increased financial and economic fragility.

We conclude that, contrary to the conventional wisdom, infrastructure investments do not typically lead to economic growth. Overinvesting in underperforming projects instead leads to economic and financial fragility. For China, we find that poorly managed infrastructure investments are a main explanation of surfacing economic and financial problems. We predict that, unless China shifts to a lower level of higher-quality infrastructure investments, the country is headed for an infrastructure-led national financial and economic crisis, which—due to China's prominent role in the world economy—is likely to also become a crisis internationally. China is not a model to follow for other economies—emerging or developed—as regards infrastructure investing, but a model to avoid.